# Photoluminescence spectroscopy of hybridized exciton-polariton coupling in silicon nanodisks with nonradiative anapole radiations

Burak Gerislioglu[1] and Arash Ahmadivand[2]

[1]Department of Physics & Astronomy, [2]Department of Electrical & Computer Engineering,
6100 Main St, Rice University, Houston, Texas 77005, United States
e-mail address: aahmadiv@rice.edu

Semiconductor nanoparticles and nanostructures in the strong coupling regime exhibit an intriguing energy scale in the optical frequencies, which is specified by the Rabi splitting between the upper and lower exciton-polariton states. Technically, exciton-polaritons are part-light, part-matter quasiparticles that arise from the strong interaction of excitonic substances and photonic platforms. In this work, using full-wave numerical and theoretical studies, we showed the emergence of strong light-matter coupling between the nonradiating anapole states from an individual semiconductor nanodisk coupled to a J-aggregate fluorescent dye molecule resonating in the visible spectrum. By demonstrating the physical mechanism behind the observed energy splitting for various Lorentzian linewidth of excitonic material, we theoretically confirmed the obtained spectral responses by conducting photoluminescence spectroscopy analysis. The coupling of anapole resonances in semiconductor nanoparticles with excitonic levels can propose interesting possibilities for the control of directional light scattering in the strong coupling limit, and the dynamic tuning of deep-subwavelength light-matter coupled states by external stimuli.

Strong light-matter coupling at the subwavelength scales constitute a fundamental field of study in nanophotonics, since it facilitates new paths toward discovering modern physical principles and promising applications in active optoelectronics and quantum optics [1-6]. Of particular interest is the high-refractive index dielectric nanoparticles where both the magnetic and electric dipole moments (Mie resonances) can be excited *via* hitting of light with the frequency below or near the bandgap frequency of the material [7]. These Mie-type resonances are accompanied by a characteristic resonant dispersion in the linear response regime, ruling by the transition amplitude of an electron from the valence band to the conduction band. In particular, this process results in the formation of an *excitonic state*, with specific binding energies [8]. In multicomponent systems, the strong and coherent coupling between the arising excitonic states (from organic and dye molecules, transition metal dichalcogenide semiconductor monolayers) and electric/magnetic states in subwavelength dielectric nanoparticles gives rise to the creation of new hybridized eigenstates, known as cavity-exciton polaritons, in the coupled systems with intriguing optical properties [8,9]. The strength of this interference is proportional to the mode volume ($V$), dipole moment of the quantum emitter ($\mu_e$), and number of emitters ($N$) that are contributing to the interaction with the nanocavity as: $g \propto \mu_e (N/V)^{1/2}$.

The formation of pronounced cavity-exciton polaritons through the strong coupling in nanophotonic and plasmonic systems has been reported in diverse platforms from single nanoparticles to the engineered meta-atoms [8-12]. This technology has offered many possibilities for exploring exotic phenomena such as active control of stimulated and spontaneous emissions [13], as well as nonradiative energy transfer process [14], and Bose-Einstein condensation [15]. In the modern photonics regime, instead, hybridized excitonic states have enabled the implementation of several important processes and tools, such as nonlinear harmonic signal generation [16-19], all-optical transistors [20], ultralow-threshold lasing [21], quantum chemistry [22], and luminescence lifetime enhancement [23]. In addition, this approach has led to the rise of single-photon optoelectronic instruments [24,25].

Pioneering efforts in exciton-polariton coupling and Rabi oscillation studies date back to the past decade with the demonstration of the interactions between excitons and classical resonances in strongly coupled systems. Recently, this concept has been generalized to the plasmon-exciton interaction in Fano-resonant architectures coupled to an atomically thin $WS_2$ layer [26] and $CsPbBr_3$ nanocrystals [27]. Driven by the ongoing race to boost both splitting energy and coupling strength, researchers have focused on the hybridized eigenstates in well-engineered cavities with unconventional spectral features. Newly, plexcitonic states have been reported in organic molecules-enhanced meta-atoms with spinning charge-current excitations [28], known as *toroidal modes* [29-44]. The ability to tightly confine the electromagnetic fields in a tiny spot by toroidal excitations enabled giant Rabi oscillations. Moreover, it is demonstrated that the excitation of a dynamic toroidal dipole with high-quality (high-$Q$) narrow lineshape, low-mode volume, and ultrafast relaxation time allows for realizing robust coupling between excitons and plasmons.

The other counterpart of a toroidal moments in meta-optics is a nonradiating *anapole* mode, which is a direct yield of the destructive interference between collocated and coaligned toroidal and electric dipole moments [45,46]. Indeed, the identical radiation pattern of both toroidal and electric dipoles leads to the net zero contribution to the far-field radiation by dynamic anapole resonances [47,48]. The excitation of such a dark-type moment in dielectric nanoparticles provides a promising path for an efficient control of light–matter interactions. This has led to the emergence of practical applications of anapole-resonant quasi-planar nanophotonic metasurfaces in the emission of nonlinear harmonic signals [49-52] and the generation of ultrafast pulses [53]. Although detailed studies have been conducted to understand the excitation mechanisms, intrinsic properties, and detection methods of anapole resonances in both single-particle and flatland metasystems, the density of states (DOS) and luminescence properties of these nontrivial excitations in the direct interference with excitonic states remain unexplored.

In this work, we report on the resonance coupling in nanophotonic heterostructures consisting of silicon (Si) nanodisks covered by fluorescent dye molecules (J-aggregate) with excitonic properties. By employing both theoretical and numerical analyses, we probe the coupling between the optically induced anapole excitations and excitonic levels. We show that the strong interference between these energy states gives rise to the formation of hybridized modes and resulting in an anticrossing behavior on the energy diagram. To validate the resonance coupling and mode splitting in the J-aggregate-covered Si nanodisk, we obtained the generated photoluminescence (PL) spectra through the calculation of the photonic density of states (PDOS), and the corresponding quantum-yield (QY) is quantified. This enabled us to precisely define the existence of an anapole excitation and its interference with excitons from the dye molecules.

To begin with, we systematically investigate the spectral response of a Si nanodisk in an uncoupled regime, by inspecting the scattering of light from a Si nanodisk located on a SiO$_2$ substrate with judiciously defined geometries. The excitation mechanism of a nonradiating anapole moment in dielectric nanoparticle was discussed in details by Miroshnichenko et al. [45], hence, we briefly explain the electromagnetic responses of our proposed design. In Figure 1a, the computed scattered electric field ($E_{\text{scat}}(\mathbf{r})$) to the far-field zone (Eq. S1 in Supplemental Materials) is shown for the p-polarized beam excitation, demonstrating the formation of a distinct anapole mode around $\lambda$~700 nm. It should be noted that the position of the anapole mode is adjusted in such a way that to spectrally overlap with the J-aggregate exciton extreme [54]. Our studies for the induced electric energy ($W_E$) inside the dielectric nanodisk exhibit the existence of the peak of this component at the anapole moment wavelength. The electric-field (E-field) intensity distribution maps in both top and cross-sectional views for the Si nanodisk at the resonant mode position are illustrated in Figures 1b and 1c, respectively. The field distribution in these graphs are consistent with the predicted surface current density maps (J) across the nanodisk at the anapole wavelength. Additionally, the magnetic-field enhancement profiles in both top and side-views are presented in Figure S1, showing the formation of strong contours of the disk corresponding to the position of the maximized surface current density. To further validate the excitation of the anapole mode, we conducted multipole decomposition analysis, in which the Cartesian moments of a dielectric nanodisk are computed using standard expansion relations (Eq. S2 in Supplemental Material). Figure S2 in Supplemental Materials explicitly illustrates the scattered power by individual multipoles, confirming far-field cancelation of the electric and toroidal dipole modes, leading to the excitation of the anapole mode.

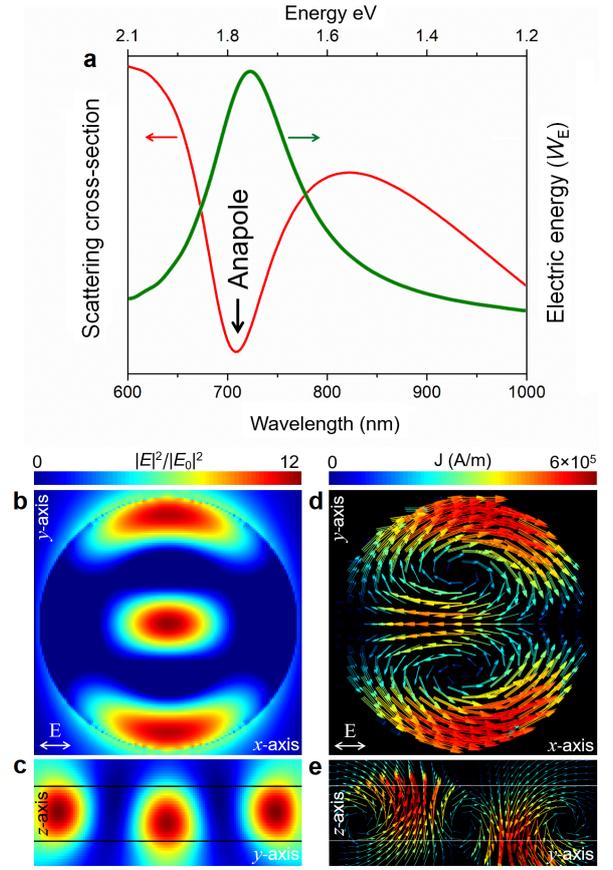

**Figure 1. Electromagnetic responses of a Si nanodisk with the diameter and thickness of 360 nm and 50 nm, respectively.** (a) Simulated scattering spectra ($E_{\text{scat}}(\mathbf{r})$) and electric energy inside the nanodisk ($W_E = n^2 \iiint |E|^2 / 2 \, dV$) under incident linearly p-polarized light. (b) Top and (c) cross-sectional view of distribution of the electric-field intensity ($|E|^2/|E_0|^2$). (d) Top and (e) cross-sectional surface current (J) distribution profiles.

In continue, we analyze the variations in the spectral response of the anapole-resonant nanodisk in the presence of J-aggregate coating layer. The spectra for the cyanine J-aggregate (5,6-dichloro-2-[3-[5,6- dichloro-1-ethyl-3-(4-sulfobutyl)-benzimidazol-2-ylidene]- propenyl]-1-ethyl-3-(4-sulfobutyl)-benzimidazolium hydroxide), inner sodium salt (TDBC) (see Figure 2a) molecular layer on a glass substrate is plotted in Figure S3 (Supplemental Materials), which possesses strong excitons in the visible bandwidth, around $\lambda \sim 700$ nm [54]. This dye molecule generates robust excitonic PL spectra [55], and in the strong coupling limit, excitons allow for significant resonance splitting [56]: $\Omega = \left(\left(4g^2 - \left[(\gamma_{opt} - \gamma_0)/2\right]^2\right)\right)^{1/2}$, where $\gamma_{opt}$ and $\gamma_0$ are the dissipation rates of uncoupled polaritons and excitons, respectively, and $g$ is the coupling strength, given by: $g = \mu_e \left(N \omega / 2\varepsilon\varepsilon_0 V\right)^{1/2}$, where $N$ is the number of emitters characterized by the transition dipole moment ($\mu_e$), $\omega$ is the transition frequency, $\varepsilon_0$ is the permittivity of vacuum, and $V$ is the mode volume ($V = \int \varepsilon E^2 dV / \max\{\varepsilon E^2\}$). It is important to note that the lowest electronic transition in TDBC is mainly described by a large transition dipole of the order of $|d| = \langle \xi | q.r | \vartheta \rangle \approx 10$ Debye (or 2.08 $e$ Å, $e$ is the electron charge), in which $\vartheta$ and $\xi$ are the ground and the first excited electronic states of the molecule, as depicted in Figure 2b. This panel displays the interaction of a single molecule with light, in which the TDBC molecule, firstly, in the ground electronic and vibrational states, is excited by absorbing light (upward arrow). Along the absorption process of the incoming photon, the positions of the nuclei remain unchanged. Furthermore, the light absorption induces specific molecular vibrations. Owing to the interference with the media, the molecular vibrations equilibrate and the molecule relaxes to the bottom of the excited electronic state (downward arrow). Consequently, the molecule relaxes down to the ground electronic state by emitting a photon correlated with the fluorescence radiation. It should be underlined that the emission from the J-aggregate monomer is a mirror–image of the absorption feature, where the peak is pushed toward the lower energies (Stokes shift) [57].

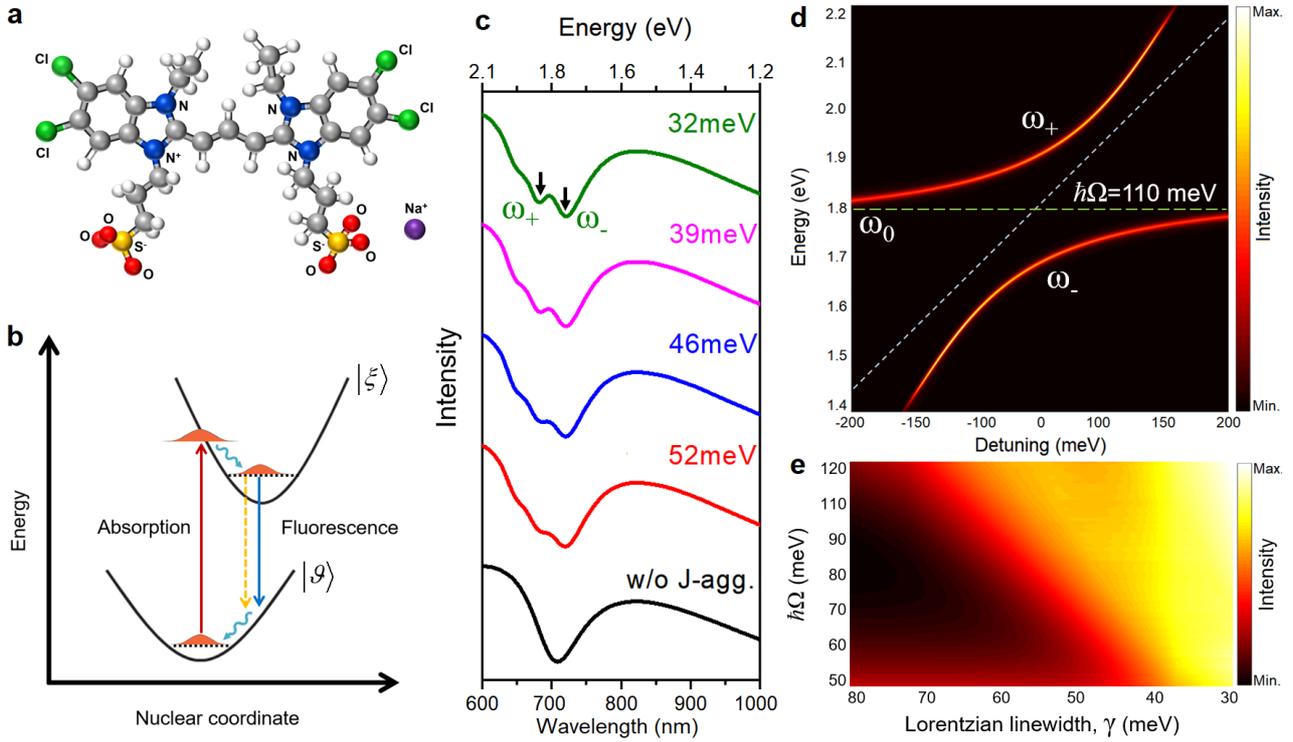

**Figure 2. The demonstration of strong exciton-polariton coupling on a single Si nanodisk covered with a TDBC molecule layer.**
(a) Molecular structure of the TDBC. (b) Excitation/Fluorescence processes in a single J-aggregate molecule illustrated on a two-level molecular energy diagram. The parabolic surfaces correspond to the ground state $|\vartheta\rangle$ and the first excited electronic $|\xi\rangle$ state of the employed molecule. The horizontal axis indicates the displacements of nuclei from their equilibrium positions. (c) Scattering spectra of the developed heterostructure for various Lorentzian linewidth ($\gamma_0$), showing splitting of the anapole moment. (d) 2D contour plot of the calculated dispersion with higher ($\omega_+$) and lower ($\omega_-$) energy bands varied as a function of detuning. The longitudinal and oblique dashed lines indicate the polariton and exciton frequencies, respectively corresponding to their energy. (e) 2D contour plot of the anapole moment splitting as a function of Lorentzian linewidth.

In the exciton-polariton coupling regime, we further study the strong interaction between the excited energy levels by investigating the influence of the Lorentzian linewidth of the dye molecule on the splitting energy. Considering the spectrum of radiated light by the excitonic and polaritonic emitters, one can define these emissions based on Fourier transform of the autocorrelation function of the excited eigenmodes (see Eqs. S3 and S4 in Supplemental Materials). Figure 2c exhibits the splitting in the anapole resonance by varying the Lorentzian linewidth

($\gamma_0$) in the dielectric permittivity of J-aggregate molecule: $\varepsilon(\omega) = \varepsilon_\infty + \left[ f\omega_0^2 / (\omega_0^2 - \omega^2 - i\gamma_0\omega) \right]$, where $\varepsilon_\infty$=2.5, $f$ is the reduced oscillator strength, and $\hbar\omega$ is the exciton transition energy. By tuning the parameters for the conventional one-oscillator Lorentzian model above, and incorporating the dye molecule, we explicitly observe the appearance of resonance splitting. The upper ($\omega_+$) and lower ($\omega_-$) frequencies corresponding to the splitting resonance are superimposed inside the panel. Obviously, the narrower the Lorentzian linewidth is, the larger the splitting energy is. The illustrated results in Figure 2c imply that instead of a quenching minimum, a scattering shoulder appears close to the exciton transition wavelength. The intensity of this spectral shoulder enhances by increasing the oscillator strength and a simultaneous splitting occurs at the dipole moment position, accompanying with the formation of two distinct eigenmodes due to the coherent energy exchange in the nanodisk. On the other hand, to better understand the splitting quality of the anapole state, we computed and mapped the 2D counter snapshot of the dispersion as a function of detuning (see anticrossing arc across the zero detuning for upper and lower bands in Figure 2d), verifying the resonance splitting of $\hbar\omega$=110 meV. Such an anticrossing trend proves the robust cavity-exciton coupling in the TDBC-covered dielectric nanodisk. Therefore, detuning of the anapole mode from the center of the exciton emission line leads to the substantial coupling strength enhancement. The influence of the Lorentzian linewidth on the splitting of the anapole moment is illustrated in Figure 2e, theoretically confirming the defined amount of the splitting. To analyze the strength of the interference between the arising eigenstates, one should evaluate the splitting value and the electromagnetic mode damping via satisfying one of the following conditions [58,59]:

$$\begin{cases} \Omega > \dfrac{\gamma_{opt} - \gamma_0}{2} \\ \Omega > \dfrac{\gamma_{opt} + \gamma_0}{2} \end{cases} \quad (1)$$

Our precise numerical simulations allowed us to determine the anapole mode bandwidth ($\gamma_{opt}$) as 150 meV. Hence, the obtained quantitative results for the polariton and exciton linewidths in the largest splitting regime, verify the fulfilment of the declared condition above for the strong coupling.

Additional insights into the coupling between arisen excitonic and polaritonic states require the knowledge of the respective DOS of photons and subsequently PL enhancement process in the developed subwavelength heterostructure. To this end, by carrying out finite-element method (FEM) studies, we straightforwardly quantified the DOS enhancement of photons and extrapolated the radiative photonic density of states (PDOS) spectra as shown in Figure 3a. Here, we ignored the PDOS in a Si nanodisk in the absence of excitonic TDBC layer. Following similar trend as seen in the scattering spectra, we observed a continuous enhancement in the PDOS response for much narrower Lorentzian linewidth. This effect can be understood clearly for $\gamma_0$=32 meV, for the projected PDOS spectra to the far-field, two shoulders associating with the excitonic and photonic states are appeared. Mathematically, the intrinsic merit of the developed exciton-polariton coupling can be further confirmed by computing the luminescence enhancement. To this end, by considering the defined PDOS response, the PL response is estimated by incorporating the probability of electron occupation of a given state in the developed platform. This probability can be described by applying the associating *T*-matrix elements. Theoretically, the emission decay rate is proportional to the PDOS that the photonic subwavelength structure exhibits for the decay. Figure 3b shows the PL intensity for the exciton-polariton coupling in the Si heterostructure.

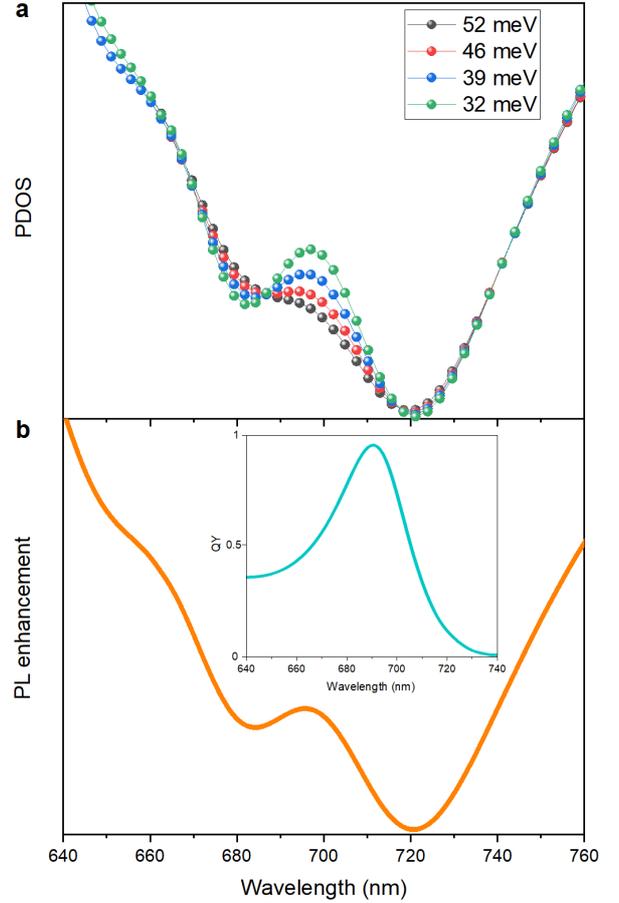

**Figure 3. DOS properties and PL response of a TDBC-covered Si Nanodisk.** (a) Calculated PDOS spectra and (b) PL enhancement for the dielectric heterostructure. Inset is the normalized QY profile for the exciton-polariton coupling process.

Ultimately, we quantified the QY of the heterostructure by dividing the number of emitted photons ($N_e$) from the system by the number of absorbed photons ($N_a$), as below [60]:

$$QY = \frac{N_e}{N_a} \qquad (2)$$

in which the number of absorbed photons in the beginning of the process was computed by: $N_a = C_a I_i / h\upsilon_e$ (where $C_a$ is the absorption cross-section, $I_i$ is the power density of the incident light, $h$ is Planck's constant, and $\upsilon_e$ is the frequency of the incident beam). It should be underlined that the intensity of beam radiated by the emitter is computed as the Fourier transform of the autocorrelation function for the excited state of the polaritonic and excitonic emitters (see Eqs. S4 and S5 in Supplemental Materials). On the other hand, we utilized the obtained data in our PL studies to estimate the number of emitted photons. The quantified results for the QY are presented as inset in Figure 3b, representing the analysis for the narrower lorentzian linewidth. Here, by narrowing the Lorentzian linewidth of the excitonic molecule, and maximizing the spectral overlap between the excitation wavelength and the anapole state, we obtained highest QY for our design.

In conclusion, we have demonstrated the exciton-polariton coupling in an anapole-resonant all-dielectric nanodisk covered with a fluorescent dye molecule. By demonstrating the excitation of a nonradiating anapole state in the proposed nanostructure, we studied the possibility of strong interaction between the polaritonic and excitonic energy levels, which resulted in the formation of substantial Rabi oscillations and splitting of the anapole moment (110 meV). The influence of the intrinsic properties of the employed TDBC molecule (*e.g.* Lorentzian linewidth) on the coupling strength as well as the Rabi oscillations was defined. The strength of the coupling was further confirmed by conducting PDOS, PL, and QY investigations. Our full-wave numerical analyses showed that the strong coupling between the arising states leads to the emergence of new hybridized levels in the heterostructure and resulting in an anticrossing behavior on the energy diagram. We envisage that the proposed mechanism can be employed as a reliable strategy for various applications including but not limited to nonlinear photonics, quantum information processing, quantum chemistry, and sensing.

## Methods

**Electromagnetic simulations.** The finite difference time domain method (FDTD, Lumerical 2019) was used for simulating the scattering and absorption spectra, E-field maps, and multipole decomposition of the anapole-resonant structure. The simulation workplace was surrounded by perfectly matched layers (PML) with 16 absorptive layers. The spatial gridding sizes was set to 1 nm is all three axes, and the time step was fixed to 0.01 fs to satisfy the Courant stability condition for all frequencies and substances [61-70]. The refractive index of both $SiO_2$ and Si was obtained from experimentally defined values by Palik [71]. We also used finite-element method (FEM, COMSOL Multiphysics 5.4) and MATLAB simulations to extract and extrapolate the PDOS, PL, and QY spectra.

# Supplemental Materials to:

# Photoluminescence spectroscopy of hybridized exciton-polariton coupling in silicon nanodisks with nonradiative anapole radiations

*Burak Gerislioglu,[1] and Arash Ahmadivand[2]\**

[1]*Department of Physics and Astronomy, Rice University, 6100 Main St, Houston, Texas 77005, United States*

[2]*Department of Electrical and Computer Engineering, Rice University, 6100 Main St, Houston, Texas 77005, United States*

*\*aahmadiv@rice.edu*

**S1. Magnetic-field enhancement**

**S2. Scattered electric-field and multipole decomposition**

**S3. Surface current density calculations**

**S4. The scattering spectra for a J-aggregate layer**

**S5. The emitted spectra from polaritonic and excitonic states**

## 1. Magnetic-field enhancement

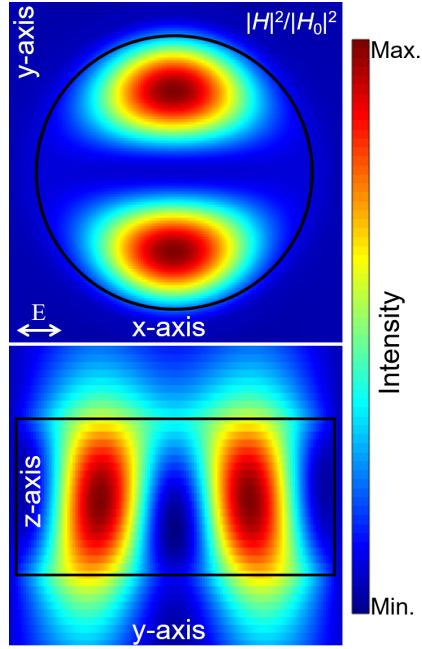

**Figure S1.** Top and cross-sectional images of magnetic-field enhancement in the Si nanodisk at the anapole wavelength.

## 2. Scattered electric-field and multipole decomposition

The scattered electric field from the Si nanodisk in the far-field zone is given by [1]:

$$E_{scat}(r) = \frac{k_0^2}{4\pi\varepsilon_0 r}\left\{[\mathbf{n}\times(\mathbf{p}\times\mathbf{n})] + \frac{ik_d}{6}\left[\mathbf{n}\times(\mathbf{n}\times\hat{\mathbf{Q}}(\mathbf{r})\times\mathbf{n})\right] + \frac{1}{\upsilon_d}[\mathbf{m}(\mathbf{r})\times\mathbf{n}] + \frac{ik_d}{2\upsilon_d}\left[\mathbf{n}\times\hat{\mathbf{M}}(\mathbf{r})\times\mathbf{n}\right] + \frac{ick_d}{\upsilon_d}\left[\mathbf{n}\times(\hat{\mathbf{T}}(\mathbf{r})\times\mathbf{n})\right]\right\}$$
$$\exp(ik_d(r-\mathbf{nr}))$$

(S1)

where $k_0$ and $k_d$, are the wavenumber in vacuum and in the medium, respectively. $c$ is the velocity of light in vacuum, and $\upsilon_d$ is the light pulse velocity in the medium around the emitter. In addition, electric dipole (**p**), electric quadrupole (**Q(r)**), magnetic dipole (**m(r)**), magnetic quadrupole (**M(r)**), and toroidal dipole moments (**T(r)**) are the Cartesian multipole moments of the scattering object that can be employed to validate the excitation of anapole state. To this end, we carried out multipole decomposition analysis, in which the Cartesian modes of a Si nanodisk are extracted using standard and general expansion relations [1,2]:

$$\mathbf{p} = \sum_{l=1}^{N} \mathbf{p}_l$$

$$\hat{Q}(\mathbf{r}) = 3\sum_{l=1}^{N}\left(\left[(\mathbf{r}-\mathbf{r}_l)\otimes\mathbf{p}_l\right]+\left[\mathbf{p}_l\otimes(\mathbf{r}-\mathbf{r}_l)\right]\right)$$

$$\mathbf{m}(\mathbf{r}) = \frac{-i\omega}{2}\sum_{l=1}^{N}\left((\mathbf{r}-\mathbf{r}_l)\times\mathbf{p}_l\right) \quad \text{(S2)}$$

$$\hat{M}(\mathbf{r}) = \frac{-i\omega}{3}\sum_{l=1}^{N}\left\{\left[(\mathbf{r}-\mathbf{r}_l)\times\mathbf{p}_l\right]\otimes(\mathbf{r}-\mathbf{r}_l)+(\mathbf{r}-\mathbf{r}_l)\otimes\left[(\mathbf{r}-\mathbf{r}_l)\times\mathbf{p}_l\right]\right\}$$

$$\mathbf{T}(\mathbf{r}) = \frac{-i\omega}{10c}\sum_{l=1}^{N}\left\{\left[(\mathbf{r}-\mathbf{r}_l).\mathbf{p}_l\right](\mathbf{r}-\mathbf{r}_l)-2(\mathbf{r}-\mathbf{r}_l)^2\mathbf{p}_l\right\}$$

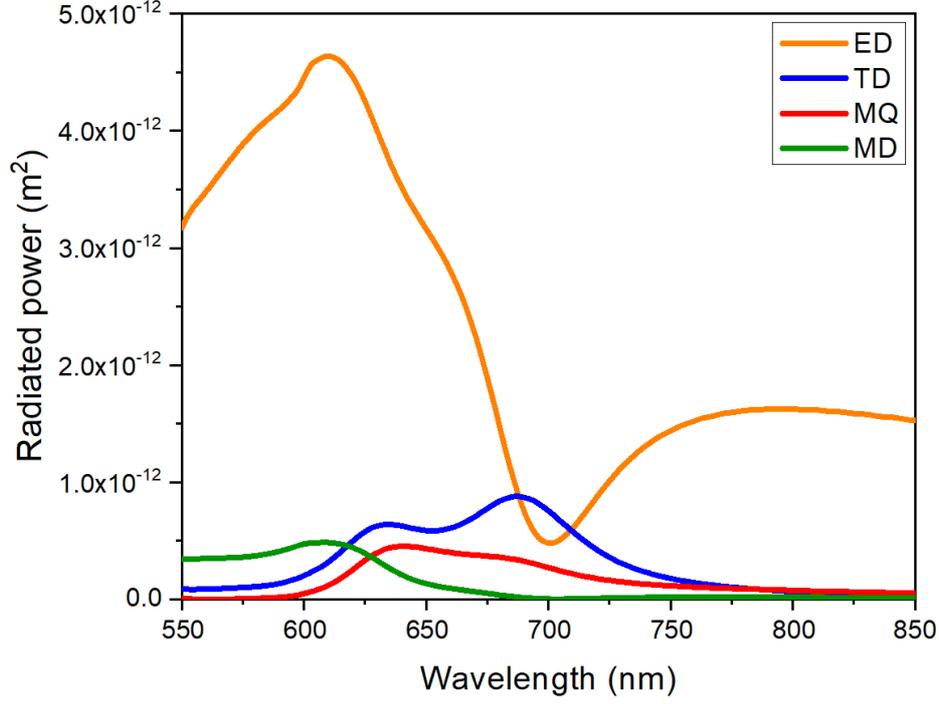

**Figure S2.** Calculated scattered power for individual electromagnetic multipoles induced in Si nanodisk by the incident radiation. ED: electric dipole, MD: magnetic dipole, MQ: magnetic quadrupole, TD: toroidal dipole.

Fig. S1 represents the scattered power by individual multipoles, validating far-field cancelation of electric and toroidal dipole modes, and leading to the excitation of an anapole mode. Under *p*-polarized plane wave excitation, clearly, both magnetic dipole and electric quadrupole moments suppress dramatically, and conversely, magnetic quadrupolar mode enhances in the far-field. Technically, this implies that the dominant behavior of magnetic quadrupole in Si disk can be perceived from the structure of the toroidal dipole moment.

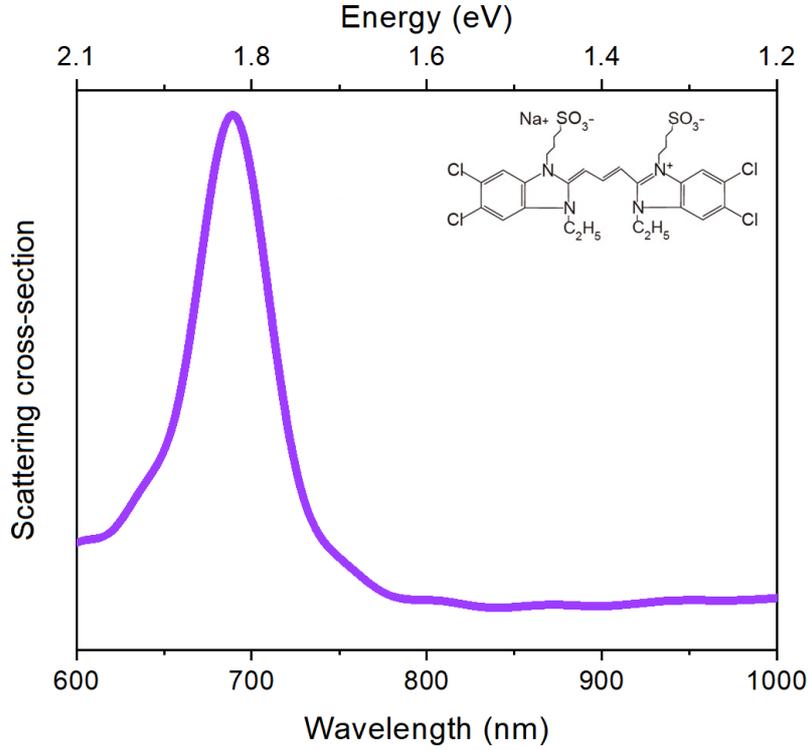

**Figure S3.** Scattering spectrum of a TDBC film coated onto a silica substrate. The inset is the molecular structure of the TDBC.

## 3. Surface current density calculations

The current density profile was determined by applying a divergence current analysis module based on implementing the effective permittivity of the system ($D = \varepsilon_e E$, where $\varepsilon_e$ is the effective permittivity of the media). Therefore, the corresponding current density can be explained by using: $J = -i\omega(\varepsilon_e - \varepsilon_0)E$, in which $\varepsilon_0$ is the permittivity of vacuum. In this way, one can rewrite the equation for the divergence current as a function of time and frequency as following: $D(\omega) = \varepsilon_0 \varepsilon_r(\omega, dt) E(\omega)$.

## 4. The scattering spectra for a J-aggregate layer

Numerically defined scattering spectra of a cyanine dye molecule, 5,6-dichloro-2-[3-[5,6- dichloro-1-ethyl-3-(4-sulfobutyl)-benzimidazol-2-ylidene]-propenyl]-1-ethyl-3-(4-sulfobutyl)-benzimidazolium hydroxide, inner sodium salt (TDBC) on a silica substrate is plotted in Figure S3.

## 5. The emitted spectra from polaritonic and excitonic states

The spectrum of light radiated by the emitter is calculated as the Fourier transform of the autocorrelation function for the excited state of the polaritonic and excitonic emitters [3,4]:

$$I_{exciton}(\omega) = \frac{\gamma_0}{2\pi} \left| \frac{\frac{1}{2\gamma_p - i(\omega - \omega_p)}}{\left[\frac{1}{4(\gamma_p + \gamma_0)} + \frac{i}{(\omega_p - \omega_0) - i(\omega - \omega_p)}\right]^2 + \Omega_R^2} \right|^2 \quad (S3)$$

$$I_{polariton}(\omega) = \frac{\gamma_p}{\pi} \left| \frac{\frac{g}{2}}{\left[\frac{1}{4(\gamma_p + \gamma_0)} + \frac{i}{(\omega_p - \omega_0) - i(\omega - \omega_p)}\right]^2 + \Omega_R^2} \right|^2 \quad (S4)$$

in which,

$$\Omega_R = \frac{1}{2}\left(g^2 + (\omega_p - \omega_0)^2 - \frac{1}{4}(\omega_p - \omega_0)^2\right)^{1/2} \quad (S5)$$